\documentclass[11pt]{article}

\usepackage[utf8]{inputenc}

\usepackage{fullpage}
\usepackage{endnotes}

\usepackage{amsmath}
\usepackage{amssymb}
\usepackage{endnotes}

\usepackage[hypertexnames=false]{hyperref}

\usepackage{ifpdf}

\ifpdf
\usepackage[pdftex]{graphicx}
\else
\usepackage{graphicx}
\fi
\title{\textbf{P.A.M. Dirac \\and the Discovery of Quantum Mechanics}}
\author{Kurt Gottfried\footnote{Electronic mail: kg13@cornell.edu}\\Laboratory for Elementary Particle Physics\\Cornell University, Ithaca New York 14853}

\date{}

\begin{document}

\ifpdf
\DeclareGraphicsExtensions{.pdf, .jpg, .tif}
\else
\DeclareGraphicsExtensions{.eps, .jpg}
\fi

\maketitle

% \begin{abstract}

Dirac's contributions to the discovery of non-relativist quantum mechanics and quantum electrodynamics, prior to his discovery of the relativistic wave equation, are described. 
% \end{abstract}

\section{Introduction}

Dirac's most famous contributions to science, the Dirac equation 
and the prediction of anti-matter, are known to all physicists. But as I have learned, 
many today are unaware of how crucial Dirac's earlier contributions were -- that he 
played a key role in the discovery and development of non-relativistic quantum mechanics, 
and that the formulation of quantum electrodynamics is almost entirely due to him. I therefore restrict myself here to his work prior to his discovery of the Dirac equation in 1928.\endnote{ This article is an expanded version of   K. Gottfried, ``P.A.M. Dirac and the Discovery of Quantum Mechanics,"  arXiv:quant-ph/0302041v1, February 5, 2003, written in connection with the centenary of Dirac's birth in 2002; see also K. Gottfried, ``Matter all in the mind,'' Nature \textbf{419}, 117 (2002).}

 Dirac was one of the 
great theoretical physicist of all times. Among the founders of `modern' 
theoretical physics, his stature is comparable to that of Bohr and Heisenberg, and surpassed only by Einstein. 
Dirac had an astounding physical intuition combined with the ability to invent new mathematics 
to create new physics. His greatest papers are, for long stretches, 
argued with inexorable logic, but at crucial points there are critical, illogical jumps. Among the inventors of quantum mechanics, only deBroglie, Heisenberg, 
Schr\"odinger and Dirac wrote breakthrough papers that have such brilliantly 
successful long jumps. Dirac was also a great stylist. In my view his book \textit{The Principles of Quantum Mechanics}  belongs to the 
great literature of the 20th Century; it  has an austere tone that reminds me of Kafka.\endnote{The first 1930 edition is considerably more austere than the later editions.} When we speak or write quantum 
mechanics we use a language that owes a great deal to Dirac.

First, a few words about Dirac's life.\endnote{See especially the recent, detailed, non-technical biography of Dirac by Graham Farmelo, ``The Strangest Man" (Basic Books, New York 2009).}\,\endnote{ H.\ Kragh, \textit{DIRAC, A Scientific Biography} (Cambridge University Press, 1990)} He was born on August 8, 1902, in Bristol. 
Dirac's father, who was Swiss, was a domestic tyrant: 
he forced his children to speak to him only in French, forbade most social contacts, and 
compelled them to pursue studies in which they were not interested. Paul's older 
brother committed suicide and Paul was exceptionally introverted and reclusive even by 
the standards of theoretical physics. Dirac's difficult relationship with his father is evident 
from the fact that he invited only his mother to Stockholm when he was awareded 
the 1933 Nobel Prize. 

Dirac graduated in electrical engineering from Bristol at the age of 19. He won a 
scholarship to Cambridge but could not afford to accept it. He stayed on in his parent's 
house, took a second degree in mathematics at Bristol, and then won a better scholarship 
which allowed him to move to Cambridge in 1923 where he became a research student of 
R.H. Fowler, a prominent theorist. Dirac was to stay at Cambridge until his retirement in 
1969. He then accepted a professorship at Florida State University, and died in Florida in 1984.

Reminiscences by two of Dirac's colleagues offer glimpses of his personality. Nevill Mott, a Cambridge colleague during Dirac's rise to fame, observed that\endnote{N.F.\ Mott, \textit{A Life in Science} (Taylor and Francis, London 1986)} 
``Dirac is rather like one’s idea of Gandhi. He is quite indifferent to cold, 
	discomfort, food, \ldots He is quite incapable of pretending to think anything that 
	he did not really think. In the age of Galileo he would have been a very contented 
	martyr." 
In a conversation I had with Niels Bohr in 1959 about superconductivity, he suddenly remarked that \endnote{Farmelo's title (Ref. 3) stems from this anecdote, which appeared as footnote 7 in ref. 1. } ``Dirac was the strangest man who ever visited my institute." I asked why, and  Bohr responded that ``during one of Dirac's visits I asked him what he was doing. He replied that he was trying to take the square-root of a matrix, and I thought to myself what a strange thing for such a brilliant man to be doing. Not long afterwards the proof sheets of his article on \textit{the} equation arrived, and I saw he had not even told me that he had been trying to take the square root of the unit matrix!" Dirac was so shy that he wanted to refuse the Nobel Prize, but changed his mind after Rutherford warned him that turning it down would 
produce far more publicity.

Although Dirac focused on physics obsessively, he loved to travel and to walk in the 
mountains, where he displayed exceptional endurance. He travelled around the world three 
times, first in 1929 in the company of Heisenberg  from Yellowstone to Japan. They were both 
becoming quite famous by then, and the press wanted an interview when their boat docked in Japan. Heisenberg knowing how shy his colleague was, told the reporters Dirac was 
unavailable even though he was standing right beside him.

Because I only deal here with two years of Dirac's career, a list of the highlights in 
his scientific  life is in order:\endnote{For a comprehensive account see  R.H.\ Dalitz and R.E.\ Peierls, \textit{Biographical Memoirs of Fellows of the Royal Society}\textbf{ 32}, 138--185 (1986).}	
\begin{itemize}
	\item November 1925 -- reformulation of Heisenberg’s groundbreaking paper; canonical 
	quantization (overlap with Born, Heisenberg and Jordan) 
	\item August 1926 -- identical particles, symmetric and antisymmetric wave functions, Fermi-Dirac 
	statistics (overlap with Fermi and Heisenberg) 
	\item December 1926 -- transformation theory -- his favorite paper (overlap with Jordan)
	\item February 1927 -- quantum theory of radiation, emission and absorption
	\item April 1927 -- scattering of light
	\item January 1928 -- Dirac equation 
	\item December 1929 -- proposes hole theory, with proton as hole	
	\item 1930 -- first edition of \textit{The Principles of Quantum Mechanics}
	\item September 1931 -- predicts anti-matter, in same paper as magnetic monopole
	\item 1932 -- appointed Lucasian Professor of Mathematics (chair once held by Newton)
	\item 1933 -- Lagrangian in quantum mechanics (forerunner to Feynman path integral) 
	\item 1933 -- Nobel Prize -- shared with Schr\"odinger
	\item 1934 -- vacuum polarization, charge renormalization
	
\end{itemize}

For the discovery of quantum mechanics, 1925 was the watershed year.\endnote{For a collection of the critical papers (all in English) that set the stage for discovery of quantum mechanics, and the discovery papers by the G\"ottingen group and Dirac, with an extensive and  insightful commentary, see  B.L.\ van der Waerden, \textit{Sources of Quantum Mechanics}  (North Holland, Amsterdam, 1967).   } By that point 
the following facts and folklore were well established:
\begin{enumerate}
	\item The Bohr-Einstein relation between atomic energy levels $W$ and the frequencies $\omega$ of emitted or absorbed
	radiation, $W_n-W_m = \hbar \omega_{nm}$. From a classical viewpoint this is deeply mysterious 
	because the radiation should have the individual orbital frequencies, not their differences. 
	\item Bohr's Correspondence Principle: results from quantum theory reduce to classical 
	physics in the limit of large quantum numbers. 
	\item Einstein's relations between the rates for absorption, and spontaneous and stimulated emission.	
		 Spontaneous emission was really unexplained until Dirac's radiation theory in 1927.
	\item The Pauli exclusion principle and electron spin.
	\item Particle-Wave duality: advocated by Einstein for light since 1905 with ever stronger arguments, but controversial until the discovery of Compton scattering in 1923. de Broglie's 
	matter wave hypothesis of 1924 played no role in development of quantum mechanics 
	until Schr\"odinger's work in early 1926.	
\end{enumerate}

\section{Heisenberg's breakthrough}

Heisenberg's fundamental step in the summer of 1925 was to discard the classical concepts of 
\textit{kinematics}, not just dynamics. Here is an excerpt from an opening paragraph of his paper:\endnote{W. Heisenberg, ``\"Uber quantentheoretische Umdeutung kinematischer und mechanischer Beziehungen," Z. Phys. \textbf{33}, 879 - 893 (1925), received July 29, 1925. }
\begin{quote}
	``It has become the practice to characterize [the] failure of the quantum-theoretic 
	rules as deviations from classical mechanics \ldots This characterization has, however, little meaning 
	when one realizes that the Einstein-Bohr frequency condition already represents 
	such a complete departure from classical mechanics, or rather from the kinematics underlying this mechanics \ldots the validity of classical mechanics simply cannot be maintained \ldots it seems sensible to 
	discard all hope of observing \ldots the position and period of [atomic] electrons, 
	and to concede that the partial agreement of the quantum rules with experience 
	is more or less fortuitous. [We here] try to establish a theoretical quantum 
	mechanics,\endnote{The term `quantum mechanics" is due to Max Born.} analogous to classical mechanics, but in which only relations between 
	observable quantities occur.  One can regard the [Bohr-Einstein] frequency condition and the dispersion theory of Kramers, [Heisenberg and Born] \ldots as the most important first steps towards such a quantum-theoretical mechanics.  In this paper we shall seek to establish some new quantum-mechanical relations and apply them to  \ldots problems involving one degree of freedom."
\end{quote}
He then turns to radiation from a bound electron, and points out that non-linear expressions in terms of the velocity and position appear in quadrupole and higher order multipole radiation, and goes on to write

\begin{quote}

 `` One may inquire about the form these higher order terms would assume in quantum theory. \ldots This point  \ldots is of a purely kinematic nature. \ldots We may pose the question in its simplest form thus: 
	If instead of a classical quantity $x(t)$ we have a quantum theoretic quantity, 
	what quantum theoretic quantity will appear in place of $[x(t)]^2$?"
\end{quote}
This surprising question was motivated not only by the preceding remark about multipole radiation, but even more so because he had long puzzled how to treat the 
anharmonic oscillator in quantum theory so that the energy values would satisfy the Einstein-Bohr frequency condition (which is trivially satisfied by the harmonic oscillator where the transition frequencies are just multiples of the basic frequency). 

This ground-breaking paper is notoriously difficult to follow because it contains what seem to be ``magical" steps.  Even van der Waerden, a powerful mathematician who 
made significant contributions to quantum mechanics, admits to not being able to follow at critical points although he had 
the benefit of interviews with Heisenberg. In 2004 however, Aitchison, MacManus and Snyder published a detailed and plausible argument that fills in the missing  links in Heisenberg's paper.\endnote{I. J.R. Aitchison, D. A. MacManus and T. N. Snyder, ``Understanding Heisenberg's ``magical'' paper of July 1925," A. J. Phys. \textbf{72}, 1370 - 1379 (2004). } I will, therefore, confine myself to a very thin sketch of Heisenberg's paper.

Given that Heisenberg had decided that the classical position was an unobservable and  meaningless concept in the atomic domain,\endnote{This turned out to be a very fruitful misconception: in quantum mechanics the position is an observable! This illustrates the fact that the interpretation of the theory developed much more slowly  than its mathematical machinery.  In particular, the uncertainty principle was only formulated by Heisenberg in March 1927.} he had to propose an observable `quantum-theoretic' replacement.  Motivated by the recent work of his colleagues Born and Kramers on radiation processes, he chose  \textit{the whole set} of radiative transition amplitudes, whose squares had been identified by them as proportional to transition probabilities, and thus observable.  These amplitudes are what would subsequently be called (apart from a trivial factor) the matrix elements $x_{nm}$ of the position operator, with the time dependence $\exp i\omega_{nm}t$. This was a huge step, of course, replacing the numerical function of classical mechanics by an infinte array involving all the states of atom.  He then addressed his own question: what is the expression for $x^2$ such  that its time dependence also satifies the Bohr-Einstein condition? This led him rather directly to
\begin{equation}    (x^2)_{nm} =  \sum_k x_{nk} x_{km}  \end{equation}
``as an almost necessary consequence of the frequency combination rule." That this  is the law of matrix multiplication was not yet known to him!  He then points out (actually more than once) that ``a significant difficulty arises if we consider two quantities $x$ and $y$ \ldots whereas in classical theory $xy$ is always equal to $yx$, this is not necessarily the case in quantum theory [as here proposed]."

Heisenberg then turned to dynamics, and wanting to obey the Correspondence Principle, he thought ``it is very natural to take over the [classical] equation of motion." He considered the anharmonic oscillator, and with an intricate calculation he found the energies $W_n$ to  second order in the anharmonic term, and showed that they still obeyed the Bohr-Einstein rule.

\section{Dirac enters the scene}

Heisenberg gave a seminar in Cambridge in July 1925, right after he had done what 
I've just sketched. He barely mentioned this work in his talk, but afterwards said more about it to Fowler, 
who asked him to send proof sheets when they became available.  The proofs arrived at the end of August, and Fowler sent them to Dirac, who was on vacation in Bristol,
with the question ``What do you think of this? I shall be glad to hear'' scrawled atop the 
first page.

Until this time,  the cutting edge work on the quantum theory was primarily done in Germany 
and Copenhagen by a small group of superbly gifted people who were in close and continuous 
contact with each other: Born, Heisenberg, Kramers and Pauli were the leading figures, with 
Bohr as their father confessor. Although Cambridge was arguably the leading center of experimental 
physics, England had contributed little to the quantum theory. Thus Dirac was doubly 
isolated---by his own personality and by not being in the continental loop. When his first 
paper on quantum mechanics arrived in Germany it was a total surprise; Max Born later 
recalled this ``as one of the great surprises of my scientific life, for the name Dirac was 
completely unknown to me.''

Up to the summer of 1925 Dirac had published a handful of papers, but none addressing basic issues in quantum theory.\endnote{ Dirac's published papers are reproduced in \textit{The collected works of P.A.M. Dirac 1928--1948}, R.H. Dalitz (ed.), (Cambridge University Press, Cambridge, 1995).} He had closely studied relativity and Hamiltonian mechanics. Dirac's first reaction to Heisenberg's manuscript was dismissive --  ``  it needed about ten days or so before I was really able to master it. And I suddenly became convinced that this would provide the key to understanding the atom."\endnote{P.A.M. Dirac, \textit{Recollections of an Exciting Era,} in \textit{History of Twentieth Century Physics},  C. Weiner (ed.), (Academic Press, New York 1977).}   

In contrast to Heisenberg, who feared it was a serious flaw that his his `quantum-mechanical quantities' do not, in general, commute, Dirac believed that this property was of critical importance. His breakthrough came during one of his habitual long Sunday walks in the countryside -- ``the idea first came in a flash -- out of the blue \ldots [that] there seemed to be a close similarity between a Poisson bracket of two quantities and their commutator." But he only ``had some vague recollections" about Poisson brackets, and had to wait ``impatiently through the night" until the libraries opened,  whereupon  ``I looked up Poisson brackets \ldots and found that they were just what I needed."

I now point to some of the highlights of Dirac's first paper on quantum mechanics.\endnote{P.A.M.Dirac, ``The Fundamental Equations of Quantum Mechanics," Proc. Roy. Soc. A \textbf{109}, 642-653 (1925), received November 7, 1925.} At the outset, he drew the following lesson from Heisenberg's paper: ``\ldots it is not the equations of classical mechanics that are in any way at fault, but the mathematical operations by which physical results are deduced from them that require modification. \textit{All}  the information supplied by the classical theory can thus be made us of \ldots '' (The latter turned out to be something of an overstatement.)   He then turned to intrucing new  mathematical operations with sections entitled ``quantum algebra'' and ``quantum differentiation.''  To establish a relationship between the commutator $xy-yx$ of two `quantum quantities' and a Poisson bracket, Dirac considered  matrix elements $(xy)_{mn}$ for large values of $m$ and $n$ where the Corresponence Principle would say that the matrix should be close to diagonal,\endnote{It is unclear whether Dirac knew that he was dealing with matrices, not that that would have mattered to him, as he was so comfortable with abstract symbols obeying well-defined rules.} and Taylor expanded in the supposedly small quantity $m-n$. This led him to    
\begin{equation}
	(xy - yx) \to i \hbar \sum_i \left(\frac{\partial x}{\partial q_i} \frac{\partial y}{\partial p_i} - 
	\frac{\partial y}{\partial q_i}\frac{\partial x}{\partial p_i} \right) ,
\end{equation}
where the arrow means in the limit of large quantum numbers.

Dirac then promoted this approximate relationship to (emphasis in original)  ``the fundamental assumption that \textit{ the difference between the Heisenberg products of two quantum quantities is equal to $i\hbar$ times the [quantum quantity that correspons to] their Poisson bracket,}'' where `quantum quantity' was soon understood to mean `operator', or in Dirac's later terminology $q$-number or `observable.' 

The canonical comutation rules are, of course, an immediate consequence of this assumption, as well as the equation of motion for any `quantum quantitity' $x(t)$:
\begin{equation}
	i\hbar\frac{\partial}{\partial t}x(t) = x(t) H - H x(t).
\end{equation}
This is now called Heisenberg's equation of motion, but it does not appear in Heisenberg's first 
paper. Dirac showed, among other things, that (3) leads immediately to the Bohr-Einstein frequency condition.

Fowler submitted Dirac's first paper to the Royal Society on November 7, 1925, not knowing 
that Born and Jordan had submitted a paper whose most important results were identical 
some five weeks before.\endnote{M. Born und P. Jordan, ``Zur Quantenmechanik,'' Z. Phys. \textbf{34}, 858 - 888 (1925), received September 27, 1925. }  In short, Born \& Jordan, and Dirac, independently discovered canonical 
quantization, and thereby transformed Heisenberg's scheme into 
a complete theory closely related to classical Hamiltonian mechanics. 
An amazingly detailed and extensive description of the theory was first provided in the famous paper by Born, Heisenberg and Jordan completed in mid-November 1925.\endnote{M. Born, W. Heisenberg und P. Jordan, ``Zur Quantenmechanik II,'' Z. Phys. \textbf{35}, 557 - 615 (1925), received November 16, 1925.} 

In January 1926,   Schr\"odinger's first paper on wave mechanics appeared. At first 
Heisenberg, Dirac et al. were hostile to wave mechanics because they thought it gave the 
misleading impression that the classical concepts of continuity and visualizability would 
survive the revolution, whereas they believed that it was a central virtue of their abstract 
theory that it did not evoke such delusions. Soon enough---by the summer of 1926---first 
 Heisenberg  and then Dirac found that wave functions were invaluable in dealing with many 
body problems.\endnote{Heisenberg initiated an extensive correspondence with Dirac immediately after receiving Dirac's proofs of his first paper on quantum mechanics; see Ref. 14.} 
Their papers were the first to recognize that indistinguishability has 
profound consequences in quantum mechanics that have no counterpart whatsoever in classical 
physics. 

Heisenberg attacked the two-electron problem, helium, which had totally defeated the 
Old Quantum Theory. He discovered that the Pauli principle requires  two-electron wave 
functions to be antisymmetric, and that the large splittings between spin triplets and singlets 
was an electrostatic effect due to the correlations imposed on the wave functions by 
antisymmetry.\endnote{W. Heisenberg, ``Mehrk\"orperproblem und Resonanz in der Quantenmechanik,'' Z. Phys. \textbf{38} 411 - 426 (1926); received June 11, 1926.}

At about the same time, Dirac produced a general analysis of systems composed of identical particles.\endnote{P.A.M. Dirac, ``On the Theory of Quantum Mechanics,'' Proc. Roy. Soc. A \textbf{112} 661 - 677 (1926); received August 26, 1926. } 
He showed that particles obeying 
Bose-Einstein statistics must be in symmetric states while those obeying the Pauli principle 
must be in antisymmetric states. Unaware of Fermi's somewhat earlier derivation of the 
Fermi distribution (in which Fermi never mentioned wave functions and therefore antisymmetry), Dirac also 
derived the Fermi distribution.

Just a cursory look through the intricate and groundbreaking papers by the G\"ottingen team, Dirac and Schr\" odinger, written over a time of about the duration of today's courses in quantum mechanics, leaves an unforgetable impression of the intellectual power these people had at their beck and call.

\section{The birth of quantum electrodynamics}

 Dirac's manuscript on the absorption and emission of light  was received by the Royal Society on February 27, 1927.\endnote{P.A.M. Dirac, ``The Quantum Theory of 
the Emission and Absorption of Radiation, ''Proc. Roy. Soc. A \textbf{114} 243 - 265 (1927)}   
This was the birth of quantum electrodynamics.\endnote{The papers by Born, Heisenberg and Jordan (Refs.  17 \& 18) have final sections on the quantum theory of the electromagnetic field which are deleted from the English translations in van der Waerden's book, and can be found most readily in M. Born, \textit{Ausgew\" ahlte Abhandlungen,} Vol. 2 (Vandenhoeck \& Ruprecht, G\"ottingen 1963). They introduce matrices for the electric and magnetic fields satifying Maxwell's equations, but treat the atom as a prescribed classical source, not by quantum mechanical perturbation theory as did Dirac. Hence for  spontaneous emission they  had to assume the Einstein relation to absorption,  while scattering could  not be treated.} 
By then about one and a half years had passed since Heisenberg's first paper. Considering the ability of the pioneers, the speed with which they had advanced, and the centrality of the radiation problem, this was a long wait. In any modern text on quantum 
mechanics, the treatment of the simplest radiative processes is among the 
most straightforward chapters. So why did it take so long, and why were Dirac's brilliant 
contemporaries so impressed by this paper?

The reason would seem to be that the theoretical machine did not yet have enough horsepower to 
handle this problem. Those working with Schr\"odinger's equation rarely left the coordinate 
representation, and the matrix mechanicians primarily dealt with stationary states using the representation in which the 
Hamiltonian is diagonal. No one had a formulation that could handle processes in which 
the number of degrees of freedom change, and time dependent perturbation theory was hardly developed.  A more powerful formalism not so tied to classical 
mechanics was needed to describe radiation, and Dirac provided this with his transformation 
theory, sent to publication exactly two months before the radiation paper. Both of these papers were written in Copenhagen, where Dirac, for the first time, had the opportunity to interact with other quantum pioneers, especially Bohr and Heisenberg. Pascual  Jordan also published a sophisticated and elaborate discussion of transformation theory at the same time.\endnote {P. Jordan, ``\"Uber eine neue Begr\" undung der Quantenmechanik,'' Z. Phys. \textbf{40} 809 - 838 (1927), received December 18, 1926.  There were other important contributions to this development, especially by F. London; see M. Jammer, \textit{The Conceptual Development of Quantum Mechanics,} (McGraw-Hill, New York, 1966), chapter 6.}  Transformation theory was, however,  the last major contribution by Dirac to be discovered 
simultaneously by someone else. Starting with radiation theory, he led the advance into what became relativistic quantum field theory.

The paper on transformation theory\endnote{P.A.M. Dirac, ``The physical interpretation of the quantum dynamics,'' Proc. Roy.  Soc. A, \textbf{113} 621 - 641 (1927), received December 2, 1926.} was Dirac’s favorite---he often referred to it as ``my 
darling,'' not the sort of word he was in the habit of using. The paper's title is ``The physical interpretation of the quantum dynamics'' because its central goal is the generalization of the Born interpretation of the Schr\"odinger wave function to scalar products between arbitrary states.  The article also established the relationships between the various formulations of the theory -- wave mechanics, matrix mechanics, and his own abstract formulation using $q$-numbers and $c$-numbers.  As to the paper's form, the theory had never before this been expressed in so elegant, general, compact and abstract a form---the form with which we are familiar today. It is an easy read for us, but most of Dirac's contemporaries found it formidably abstract, and the style did not become popular until several decades later. After first hearing Dirac's presentation of transformation theory, Heisenberg, in a letter to Pauli on November 23, 1926 from Copenhagen,  wrote of Dirac's ``extraordinarily grandiose generalization of transformation theory.''\endnote{W. Pauli, \textit{Scientific Correspondence}, Vol.\ 1, A.\ Hermann, K.v. Meyenn \& V.\ Weisskopf (eds.) (Springer-Verlag, New York, 1979); pp. 357 - 360.}

Dirac began with  sections introducing his notation for $q$-numbers having discrete and/or continuous spectra; for dealing with the latter he introduced  his delta function:
\begin{quote}
	``\ldots of course, $\delta(x)$ is not a proper function of $x$, but can be regarded only as 
	the limit of a certain sequence of functions. All the same one can use $\delta(x)$ as 
	though it were a proper function for practically all purposes \ldots without getting 
	incorrect results. One can also use [derivatives] of $\delta(x)$ which are even \ldots less ``proper'' than $\delta(x)$ itself.''
\end{quote}

The notation $(\alpha'|\beta')$ is introduced for the element of the matrix that transforms from the representation in which the $q$-number $\alpha$ has the eigenvalue $\alpha'$ to the one where $\beta$ has the value $\beta'$. He later shows that the time-independent Schr\"odinger wave function $\psi_{E'}(q')$ is the transformation function $(E'|q')$ from the representation in which the coordinate has the value $q'$ to where the energy has the value $E'$.  Furthermore, he showed 
that if a system is in the state represented by the wave function 
\begin{equation} \psi_{\alpha'}(q') \equiv (q'|\alpha')  ,
\end{equation}
the probability that an arbitrary $q$-number $\Gamma$ will display its spectrum $\gamma$ in some range $(\gamma_1,\gamma_2$) is
\begin{equation}
	P_\psi = \int_{\gamma_1}^{\gamma_2} d\gamma \,\,\Big | \int dq' (\gamma|q') (q'|\alpha')\Big | ^2.
\end{equation}

Dirac's 1927 paper on radiation theory presented
 the first formulation of second quantization for bosons;  the first explanation of spontaneous emission from first principles; and the first derivation of the `Golden Rule' of time dependent perturbation theory.\endnote{For detailed accounts of the birth of quantum electrodynamics see 
S.S.\, Schweber, \textit{QED and The Men Who Made It} (Princeton University Press, Princeton, 1994));  A.I.\ Miller, \textit{Early Quantum Electrodynamics} (Cambridge University Press, Cambridge, 1994); and R. Jost, in \textit{Aspects of Quantum Theory}, A. Salam and E.P. Wigner (eds.) (Cambridge University Press, London, 1972). }

In the paper's introduction 
Dirac pointed out that his theory displays
\begin{quote}
	``\ldots complete harmony between the wave and light quantum description of 
	the interaction. We shall actually build the theory up from the light quantum point 
	of view, and show that the Hamiltonian transforms naturally into a form which 
	resembles that for waves.''
\end{quote}
This is an especially important illustration of Bohr's Principle of Complementarity, which 
was only formulated by Bohr later that year and first presented by him to the Como Conference in the 
fall of 1927.

 The paper then continues with  strings of now conventional steps, interrupted by several logical jumps.  He starts with the time-dependent Schr\"odinger equation with Hamiltonian $H_0+V$, and expands an arbitrary solution $\Psi(t)$ in terms of the stationary states $\psi_n$ belonging to $H_0$: 
\begin{equation}
	\Psi(t) = \sum_n a_n(t)\psi_n  ,
\end{equation}
so that the expansion coefficients satisfy
\begin{equation}	i\hbar \dot a_n = \sum_m V_{nm}a_m ;\end{equation}
we now call this the interaction representation. 

The system of interest is composed of $N$ indistinguishable particles obeying Bose statististics. This is used to justify the unconventional normalization 
\begin{equation}
	\sum_n |a_n|^2 = N \end{equation}  
obtained by multiplying each $a_n$ by $\sqrt{N}$, with the interpretation that now  $|a_n|^2$ \, ``is the probable number of particles  $N'_n$ in the state $n$.'' 

He then redefines the interaction Hamiltonian as
\begin{equation}
	\textsf{V} = \sum_{nm}a_n^* V_{nm}a_m  ,
\end{equation}
in which $a_n$ and $i\hbar a_n^*$ are canonically conjugate variables, \textit{and are still c-numbers!}  He then jumps to \textit {second quantization} by  defining the $q$-number creation and destruction operators 
\begin{align}
	b_n = a_n \exp(-iE_nt/\hbar) \hspace{2 cm} [b_n,b^\dag_m]=\delta_{nm}.
\end{align}
The Schr\"odinger equation was then written in the representation in which the operators $b_n^\dag b_n$ are diagonal with integer eigenvalues $N_n'$:
\begin{align}
	i\hbar\frac{\partial}{\partial t}\Psi(N_1' \ldots; t) = \sum_{nm} V_{nm} \sqrt{N_n'}\sqrt{N'_m + 1-\delta_{nm}\,
} \, \Psi(\ldots N'_n-1\ldots N'_m + 1\ldots ;t).
\end{align}
The total number of particles is still conserved! 

Now the last long jump.  Dirac observes that
\begin{quotation} ``the light quantum has the peculiarity that it apparently ceases to exist when it is in \ldots the zero state in which its momentum, and therefore its energy, is zero. When a light quantum is absorbed it can be considered to jump into the zero 
	state, and when one is emitted it can be considered to jump from the zero 
	state to one in which it is in physical evidence, so that it appears to have been 
	created.''
\end{quotation}
It is remarkable that although Dirac had just invented invented \textit{the mathematical description} of particle 
creation and destruction, he had not yet accepted \textit{the concept}.  In his conception the zero state -- the vacuum, contains an infinite number of light quanta, all those which have already disappeared in absorption and those that are still to appear in emission. 
This idea that the vacuum contains an infinite number of particles would be used by him 
again to invent hole theory.

 The assumption that $N'_0$ is infinite then  motivated the last jump:
\begin{align}
	N'_0 \to \infty \hspace{2cm} V_{0m} \to 0 \hspace{2cm} V_{0m}\sqrt{N'_0} \to v_m \quad \text{(finite).}
\end{align}
The interaction between the radiation field and matter was then shown to assume the familiar second-quantized form in the dipole approximation. After deriving the Golden Rule he used it to find  the 
rates for absorption and emission, and the Einstein relations between them.\endnote{Dirac devised the well-known hocus-pocus leading to a rate proportional to time. Those who have found this derivation troubling should be pleased that Heisenberg reported devoting a considerable effort to understanding what he called this trickery (`Mogelei' in German) in a letter to Pauli (Ref. 26, pp. 460 - 463.}

In retrospect, it appears to me that in creating quantum electrodynamics, Dirac started out with the radiation issue at the back of his mind, but first focused on describing a system of interacting bosons, where conservation of $N$ would be a natural (even unstated) assumption. Then, after inventing second quantization, he turned to radiation, where conservation of $N$ has to be gotten rid of, which he accomplished by imagining that the vacuum holds an infinite number of light quanta, which later paid off in hole theory and the prediction of antimatter. Of course, it is also so that until Fermi's theory of $\beta$-decay in 1934, no one thought of creating particles out of `nothing'. 

Two months after his first paper on radiation theory,  Dirac finished the rather complicated derivation of the Kramers-Heisenberg dispersion formula for scattering of light by extending his time dependent perturbation theory to second
order.\endnote{P.A.M. Dirac, ``The Quantum Theory of Dispersion,'' Proc. Roy. Soc. A \textbf{114}, 710 - 728 (1927), received April 4, 1927.}  In this paper he strips away the $N$-conserving underbrush he had to traverse in his previous paper, and provides a description of the theory that is far clearer to the modern reader, with second quantization used from the start.  He also developed, but did not publish, the theory of line width.\endnote{Dirac described his solution to Bohr in letter dated February 19, 1927, and reproduced in Schweber, Ref. 27,  pp. 31 - 32. As Schweber points out, this was three years before the well-known solution of this problem by Weisskopf and Wigner.}

Dirac published a general theory 
of collisions in non-relativistic quantum mechanics in July 1927. It was formulated  in momentum space in a style that was not to become popular until the work of Lippmann and Schwinger in 1950.  The paper includes a treatment of resonance scattering leading to an expression of the Breit-Wigner type, including a shift of the level.\endnote{P.A.M. Dirac, ``\"Uber die Quantenmechanik der Stossvorg\"ange,'' Z. Physik \textbf{44} 585 - 595 (1927), received June 28, 1927. Dirac's solution of the line breadth problem (Ref. 30)  framed the issue as a scattering problem.}

In October 1927, at the age of 25 and just two years after he first appeared on the scene, 
Dirac was the youngest participant in the famous and highly exclusive Solvay Congress 
where Bohr and Einstein began their long debate about the foundations of quantum mechanics. After a discussion among Dirac, Pauli and Heisenberg of philosophy and 
religion, in which Dirac expressed his distaste for such ponderings, Pauli, with his famous 
acerbic whit, quipped that ``Dirac’s religion is that there is no God, and Dirac is His 
Prophet.'' This was prescient --  the Dirac equation was received by the Royal Society on January 2, 1928.

\theendnotes

\end{document}